\documentclass[10pt,conference,letterpaper]{IEEEtran}
\usepackage{times,amsmath }

\usepackage{graphicx}
\usepackage{algorithm}
\usepackage{algorithmic}
\usepackage{latexsym} 
\usepackage{subfigure}
\newcommand{\DEFS}{\item[\textbf{Definitions:}]}
\title{Parallelizing Windowed Stream Joins in a Shared-Nothing Cluster}
\author{\IEEEauthorblockN{Abhirup Chakraborty}
\IEEEauthorblockA{School of Informatics and Computer Science\\
Indiana University, Bloomington, Indiana 47408\\
Email: achakrab@indiana.edu}
\and
\IEEEauthorblockN{Ajit Singh}
\IEEEauthorblockA{ Department of Electrical and Computer Engineering\\
University of Waterloo, ON, Canada N2L 3G1\\
Email: a.singh@ece.uwaterloo.ca}
}

\begin{document}
\maketitle

\begin{abstract}
The availability of large number of processing nodes in a  parallel and distributed computing environment enables sophisticated real time processing over high speed data streams, as required by many emerging applications. Sliding window stream joins are among the most important operators in a stream processing system. In this paper, we consider the issue of parallelizing a sliding window stream join operator over a shared nothing cluster. We propose a framework, based on fixed or predefined communication pattern, to distribute the join processing loads over the shared-nothing cluster. We consider various overheads  while scaling  over a large number of nodes, and propose solution methodologies to cope with the issues.   We implement the algorithm over a cluster   using a message passing  system, and present the experimental results showing the effectiveness of the join processing algorithm. 

\end{abstract}

% A category with the (minimum) three required fields
%\category{H.4}{Information Systems Applications}{Miscellaneous}
%A category including the fourth, optional field follows...
%\category{D.2.8}{Software Engineering}{Metrics}[complexity measures, performance measures]

%\terms{Delphi theory}

%\keywords{ACM proceedings, \LaTeX, text tagging} % NOT required for Proceedings

\section{Introduction}\label{sec:intro}
Data stream management systems (DSMS) emerge to support a large classes of applications, such as stock trading surveillance, network traffic monitoring, sensor data analysis, real time data warehousing, that require sophisticated processing over online data streams. The DSMS processes continuous queries (CQ)~\cite{babu01} over high-volume and time-varying data streams.The long running continuous queries differ from traditional request-response style queries over a persistent (non-streaming) database. In a CQ-system, users register queries specifying their interests over unbounded, streaming data sources. A query  engine continuously evaluates the query results with the arrival of incoming data from the sources, and delivers the unbounded, streaming outputs to the appropriate users. A core operator in a CQ-system is sliding window join among streams, called {\it window join}. A sliding window limits the scope of the join operator over a recent window, thus unblocking the join operator. Such a window join is relevant to many applications which need to correlate each incoming tuple with recently arrived tuples from the other streams~\cite{gedik05}. Such a window join is used to detect correlations among different data streams, and has many applications in video surveillance, network monitoring, sensor or environmental monitoring. 

\par The stream applications place several scalability requirements on the system. First, for high stream rates
and large window sizes, a sliding window join might consume large memory to store the tuples of the stream windows~\cite{srivas04}.  Second, as results need to be computed upon the arrival of incoming data, fast response time and high data throughput are essential. Third, some join queries such as video analysis can be CPU-intensive~\cite{gu06}. Fourth, a typical data stream management system could have numerous window join queries registered by the users.  Thus, a single server may not have enough resources to process the join queries over a high stream rate. There are two approaches to address these scalability issues: shedding loads to sacrifice result quality~\cite{tatbul03,srivas04,motwan03}, or diffusing  the workload to other machines~\cite{gu05}. We partition the streaming data over a Shared-Nothing cluster connected by high-speed networks~\cite{stoneb86}.

\par  Scalable processing of data streams over a distributed system has  been  studied by the researchers. Reference\cite{xing05} proposes a dynamic load distribution framework by partitioning the query operators across a number of processing nodes. Thus, thus approach provides coarse-grained load balancing with inter-operator parallelism. However, such an inter-operator parallelism doesn't allow a single operator to collectively use resources on multiple servers.  In~\cite{gu07}, the authors address the issue of  diffusing the join (both equijoins and non-equijoins) processing loads across a number of servers, and  provide two tuple routing strategies satisfying  {\it correlation constraints} for preserving join accuracy. The approaches have large network overhead, achieve poor load-balancing across the nodes, and in the worst case might result in overloading a master node receiving a major part of the processing load (Section~\ref{sec:relWorks}).  

\par In this paper, we consider the issue of parallelizing a window join over a shared-nothing cluster to achieve gradual scale-out by exploiting a collection of non-dedicated processing nodes. In such an environment, a  processing node can be shared by multiple applications; therefore, the need for over-provisioning for the peak load of any application is not necessary. As multiple applications or users share each node, the non-query background load and the available memory for processing queries vary on each of the nodes. Since the continuous stream join queries run indefinitely, the join operator will encounter changes in both system and workload  while processing the queries. In such an environment, intra-operator parallelism of a window join can be achieved by  partitioning the streams across the processing nodes and instantiating the window joins within every processing node that process the join over a subset of the partitions of the streams. To achieve optimal performance, the system should adjust the data stream partitioning on the fly to balance resource utilization.  

%The Flux operator~\cite{shah03} encapsulates the adaptive repartitioning, and allows intra-operator parallelism for a wide range of look-up based operators, e.g., stream window joins, group-by aggregate etc.      

%To parallelize the window join over a shared nothing cluster, we use a public domain software Message Passing Interface (MPI)~\cite{mpi-forum} that facilitates the use of such a  network of computers for cost-effective parallel computation. We observe that the existing dataflow operator Flux can be leveraged to process stream joins (equi-joins) over a shared nothing cluster by hash-partitioning the input streams,  distributing a subset of partitions to the available nodes, and adjusting the dataflow towards the nodes based on the availability of the resources within the nodes. 
\par We parallelize the window join over a shared nothing cluster  by hash-partitioning the input streams,  distributing a subset of partitions to the available nodes, and adjusting the dataflow towards the nodes based on the availability of the resources within the nodes.
%Considering the nature of communication primitives on MPI (e.g., a {\it receive} must always block if the sender is not available), we propose a framework to distribute the incoming tuples and adapt the loads across the slaves (i.e., the processing nodes). 
Considering the nature of communication primitives (e.g., receiving a packet must block the receiver if the sender is not available) within any persistent or reliable connection, e.g., Transmission Control Protocol (TCP)~\cite{stevens94}, we propose a framework to distribute the incoming tuples and adapt the loads across the slaves (i.e., the processing nodes). A slave node joins the incoming tuples with the partitions from the opposite streams using a simple nested loop join; other join algorithms based on sorting are not feasible as the temporal order of the tuples should be preserved to allow efficient tuple invalidation.  

\par With the increase in arrival rates, the size of the individual partitions within each partition-group increases. This phenomenon limits the scalability of the join algorithm: as partitions grow in size, the  CPU-time to scan the partitions and join with a new tuple also increases. To ameliorate this problem, we fine tune the partition-groups at each  processing node by dynamically adjusting the sizes of each partition. In summary, the key contributions of the paper are as follows:
\begin{enumerate}
\item We propose a technique to support fine-grained, intra-operator parallelism while executing a stream join operator over a shared-nothing cluster. The proposed technique doesn't assume all-time, any-to-any persistent communication among the participating nodes, eliminating the scalability overhead.
\item We observe a performance bottleneck in processing the window join over high stream rates, and propose a solution methodology based on the fine-tuning of the window partitions locally in each processing node. 
\item We analyze the overheads in scaling the system to a large number of nodes and propose the methodologies to optimize the scalability overheads of the system.
\item We implement the algorithm in a real system, and present experimental results showing the effectiveness of the techniques.
\end{enumerate} 

The rest of the paper is organized as follows. Section~\ref{sec:model} provides the basic concept in processing sliding window joins, and presents the system model considered in the paper. Section~\ref{sec:overview} defines the problem and provides an overview of the proposed algorithm. Section~\ref{sec:mpiJoin} describes load balancing technique in details. Section~\ref{sec:scala} considers the issue of scaling the system to a large number of nodes, and proposes techniques to reduce the system overheads (e.g., processing and communication overhead). Section~\ref{sec:exp} presents the experimental studies. Section~\ref{sec:relWorks} surveys the related work, and Section~\ref{sec:conclusion} summarizes the paper and presents future work.
   
\section{System Model}\label{sec:model}
The windowed join operator computes the join results over sliding windows of multiple streams.   For a stream $S_i$, we use $r_i$ to denote the average arrival rate in stream $S_i$. In a dynamic stream environment, this arrival rate can change over time. Each tuple $s \in S_i$ has a timestamp $s.t$ identifying the arrival time at the system. As in~\cite{babu05}, we assume that the tuples within a stream have a global ordering based on the system's clock.  We use $S[W_i]$ to  denote a sliding window  on the stream $S_i$, where $W_i$ is the window size in time units. Abusing the notation a little, we use $W_i$  to denote the window $S[W_i]$; the difference will be explicit from the context. At any time $t$, a tuple $s$ belongs to $S_i[W_i]$ if $s$ has arrived on $S_i$ within the interval $[t-W_i, t]$. The output of a sliding window equi-join $S_1[W_1] \Join \cdots \Join S_n[W_n]$  on a join attribute $A$  consists of all composite tuples $s_1,\ldots, s_n$, such that $\forall s_i\in S_i, \forall s_k\in S_k[W_k]$ $ 1\le k\le n, k\ne i $ at a time $s_i.t$, and $(s_1.A=\cdots = s_n.A)$.

\begin{figure} 
\centering
\includegraphics[width=6cm]{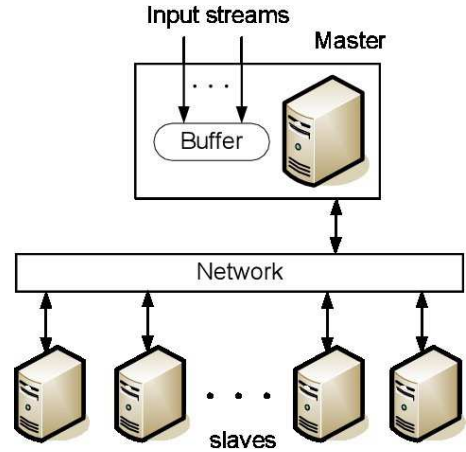}
\caption{The System Model for processing window join}\label{fig:mpi-model}
\end{figure}

\par The distributed stream processing system consists of a cluster of processing nodes connected by a high-speed network. Data streams from various external sources are pushed to a master node that serves as gateway to distribute workload across the slaves as shown in Figure~\ref{fig:mpi-model}. The join queries from the users are submitted to the master node. For a given stream join query,  the master node  selects the number of slaves to instantiate the join operator in. Moreover, the master node stores the incoming stream tuples within a buffer, and periodically sends the tuples to the slaves which carry out the actual processing.  The join results from the slaves are routed to a collector node that merges the query results and sends to the respective users. Thus, the shared-nothing stream processing system appears to an user (or client) as a unified stream processing service to serve a large number of continuous windowed join queries over  high volume data streams.   

\section{Problem Definition and Solution Approach}\label{sec:overview} 
 This section presents the problem considered in this paper, and provides various approaches to the problem and their limitations. The section ends with detailed definition of the problem. 
\par  This paper considers an operator that joins sliding windows of two streams $S_1$ and $S_2$, i.e., $W_1$ and $W_2$, respectively. For a join attribute $A$, we aim at answering the continuous join operator $W_1\Join W_2$ with condition $s_1.A=s_2.A$, for all $s_i\in W_i (i=1,2)$. The join operator over the recent windows of the streams are continuously evaluated, at different time points, with the arrival of stream tuples. Tuples in a stream are organized into $n_{part}$ partitions. Thus each window is splitted into  $n_{part}$ partitions based on a hash function $\mathcal{H}$, and denote the $j$-th partitions of a window $W_i$ as $W_i[j]$. Given the partitions of the windows, the hash join can be evaluated by simply joining the tuples within the partitions (from different joining streams) with the same partition ID, as given by \[W_1\Join W_2 =\cup_{j=1}^{n_{part}} W_1[j]\Join W_2[j]\]

\par In a distributed environment, initiating data exchange or communication by an application  at any point in time, and without any prior synchronization or predefined order, is infeasible. For example, an application receiving data over a persistent connection  must block if the sender node is not scheduled to send the data. Such a phenomenon  degrades the system performance, as an application, while blocked,  can't continue to process incoming tuples until the data from the sender is received (i.e., the receiver is unblocked). Also, asynchronous (non-blocking) communication in high speed stream applications is infeasible due to buffer overflow.

\par This paper considers the issue of computing sliding window equijoins joins  over a shared-nothing cluster forgoing any notion of persistent, any-time, all-to-all communication  pattern among the nodes. The algorithm should follow a fixed communication sequence or predefined order of data exchange among the processing nodes. The problem considered in the paper involves (a) dynamically balancing the join processing loads among the slaves, (b) determining the degree of declustering based on the communication overhead, total processing time, and the processing load  at the nodes, (c) study the tradeoff between the latencies in the output tuples and communication overhead.    
 
\section{Join processing based on Synchronous Communication}\label{sec:mpiJoin}
In this section, we introduce the  load diffusion algorithm, based on Synchronization of communication among the participating nodes, to process hash join over a shared-nothing cluster. We start with the architectural framework and present the detailed features of the system in subsequent parts of the section. 
~\cite{a}

\begin{figure} 
\centering
\includegraphics[width=6cm]{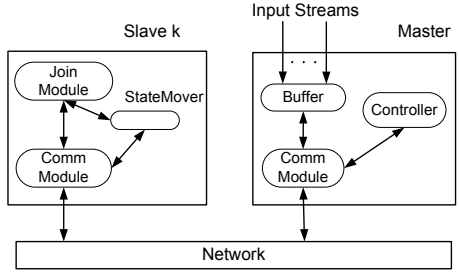}
\caption{System Architecture showing various software components at different types of nodes}\label{fig:mpi-archi}
\end{figure}

\subsection{Overview}
The join processing system consists of two categories of nodes, a master and the slaves, that communicate over  a network using  communication primitives (i.e., {\it send} and {\it receive}) over a reliable and persistent connection. The software components of the nodes are shown in Figure~\ref{fig:mpi-archi}. The master node stores the incoming stream tuples in a buffer. The slaves communicate with the master node periodically , at the end of predefined time intervals, and receive stream tuples and/or system load information. At the end of every {\it distribution epoch}, the master node sends the buffered tuples to the slave nodes; whereas, at the end of every {\it reorganization epoch}, the master node, based on the observed workloads at the slave nodes, adjusts (a) the mappings between the partition-groups and the slave nodes, and (b) the degree of declustering, i,e., the number of slaves actively participating in processing the sliding window join.  The controller module in the master node carries out the processing for reorganization. On the other hand, a slave node receives stream tuples and special instructions (e.g., move a window partition) from the master, and initiate relevant processing. Algorithm~\ref{alg:masterProcess} shows the high-level overview of the operations carried out at the  master side. 
\begin{algorithm}
\algsetup{indent=1.5em, linenosize=\small, linenodelimiter=.}
\caption{\textsc{MasterProcess}($S_i$, $SN$, $ASN$)}\label{alg:masterProcess}
\begin{algorithmic}[1]
\REQUIRE Data Streams $S_i$ $(1\le i\le 2)$, A set of slave nodes $SN$ 
\ENSURE Proper distribution of stream tuples among the active slave nodes and appropriate set of active slave nodes $ASN$  
\DEFS { \mbox{} \\ $\boldsymbol{ASN}$: set of active slave nodes\\
					$\boldsymbol{T_{clock}}$: Current system clock time\\
					$\boldsymbol{T_{dist}, T_{rep}}$: Recent distribution and repartitioning time, respectively\\
					$\boldsymbol{\delta_{dist}, \delta_{rep}}$: Distribution and repartitioning epoch, respectively\\ \mbox{} }
%\item[]
\STATE initialize $ASN$
\STATE $T_{dist}\leftarrow T{rep} \leftarrow T_{clock}$
\WHILE{(True)}
	\STATE receive tuple $s_i\in S_i$ and put in the proper mini-buffer
	\IF{$T_{clock}\ge (T_{dist}+\delta_{dist})$}
		\STATE collect tuples for each active slaves from the respective mini-buffers
		\STATE send tuples to the active slave nodes
		\STATE $T_{dist}\leftarrow T_{clock}$
		%\STATE Synchronize clocks with the active slaves  
	\ENDIF
	\IF{$T_{clock}\ge (T_{rep}+\delta_{rep})$}
		\STATE identify suppliers and consumers
		\FOR{each supplier} 
			\STATE select an unique consumer
		\ENDFOR
		\STATE Adjust the degree of declustering
		\STATE direct partition-movement information/meta-data to each $<supplier, consumer>$ pair 
		\STATE $T_{rep}\leftarrow T_{clock}$
		\STATE Synchronize clocks with the active slaves  
	\ENDIF
\ENDWHILE
\end{algorithmic}
\end{algorithm}

\begin{figure} 
\centering
\includegraphics[width=6.5cm]{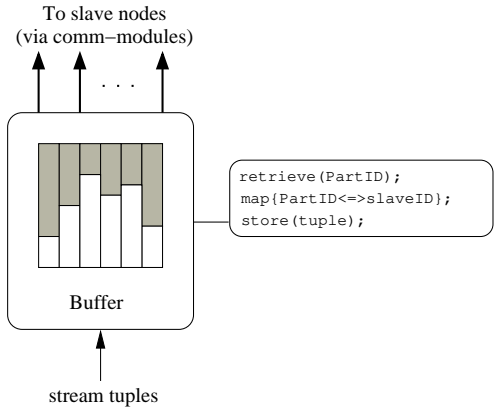}
\caption{Buffer at the master node}\label{fig:mpi-master-buffer}
\end{figure}
\subsection{Tuple distribution}
A master node maintains, for each stream, a buffer that consists of multiple "mini-buffers", one for each partition (Figure~\ref{fig:mpi-master-buffer}). The buffer keeps the mapping between the partition ID and the slave machines assigned with the partition. From such a mapping, the partitions assigned to a slave can be identified. At the end of a distribution epoch, when the slave node communicates with the master, the master drains the tuples from the "min-buffers" corresponding to the partitions assigned to each slave, and sends the merged tuples to the slave node in machine independent format. As the tuples from all the streams are merged, the merged stream, received at the client, should have enough information to map the tuples to the respective source streams. Two approaches are possible: augmenting an extra attribute, containing the stream ID, with each stream tuple; and putting special punctuation marks (which might itself be fictitious tuples) at the sequence of tuples from each stream.  

\par On the other hand, a slave node receives the tuples from the master, and stores the tuples in a stream buffer (at the join module). Once the tuple distribution phase is finished, the join module start to process the tuples from the buffer. The buffer maintained at a join module within a slave node is similar to that in the master except for the mapping information.     

%\subsection{Slave Side Architecture} % buffering, fine tuning etc..  

\subsection{Repartitioning} % load balancing, state movement etc..
The input streams are repartitioned to re-balance the works across the slave nodes. To facilitate such re-balancing of works across the slaves, we introduce a level of indirection while partitioning the input streams: instead  of having one giant partition per slave node, we instantiate numerous partitions, so that the total number of partitions is much higher than the maximum degree of declustering of the hash join~\cite{shah03,dewit92}. These numerous partitions are distributed among the operator instances at the slaves during  initialization, and those  instances are responsible for processing the corresponding inputs. The repartitioning task requires moving the window states for processing the subsequent input tuples.
\par The processing nodes in the system (master and the slaves) start the repartitioning protocol periodically at the end of an interval called reorganization epoch, which is a order of magnitude larger than the distribution epoch; a large value of the reorganization epoch is necessary to capture the long term variation in join workloads. In this protocol, the slave nodes send to the master the information about the loads. We use an average buffer occupancy metric, over the current reorganization epoch,  at a slave as the indication of the load applied  to the slave. A slave node records the  buffer size at the end of each distribution epoch within current reorganization interval and  calculate  their average. The average buffer occupancy metric is obtained by dividing the calculated average buffer size by the total buffer size (i.e., memory allotted to the buffer); we assume that the memory allocated to the buffer in every slave is the same.  Based on this average buffer occupancy ($f_{i}$) values  of the nodes, the master classifies the slaves into one of the three categories: {\it supplier}, {\it consumer}, and {\it neutral}. A {\it supplier} $i$ has an average buffer occupancy $f_{i}$ above a threshold value $Th_{sup}$, whereas a consumer is a slave with the average buffer occupancy below $Th_{con}$ ($0\le Th_{con}<Th_{sup}<1$); the rest of the nodes, which are neither supplier nor consumer, are classified as the {\it neutrals}. A supplier yields a partition-group to a consumer node, which installs the partition-group in its join module and processes the subsequent tuples arriving within the partition-group. As outlined in section~\ref{sec:scala}, in a stable system, the number of consumers is higher than that of suppliers. 

\par While reorganizing the placement of partition-groups across the nodes, each supplier yield only one randomly selected partition-group to a consumer. For each consumer, the master node selects a supplier, and sends  messages to the participating nodes  to initiate transferring the window states of the partition-groups to move. The supplier-consumer pairs can be identified by a single scan over the list of  the slave nodes. The master node makes necessary changes in the mapping between the stream partitions and the slave nodes. To transfer a state, the {\it state-mover} in a slave node (supplier) extracts the tuples from both the stream windows and the buffer (at the join module), and sends to the  consumer. The splitting information, if any, is also sent to the consumer to enable it reconstruct the fine-tuned partitions. After finishing the state movement, the participating nodes send acknowledgement to the master indicating the completion of the task, upon which  the master node transfers the pending tuples (in its buffer)  to be processed. This completes both the reorganization and  the tuple distribution  phases.  Note that the slaves not participating in the state movement receives the pending tuples before the master receives an acknowledgement from every node participating in the state movement.      
      
\subsection{Processing at the Join Module}
\begin{figure}[] 
\subfigure[]{\label{fig:mpi-join-proc} \includegraphics[width=4cm]{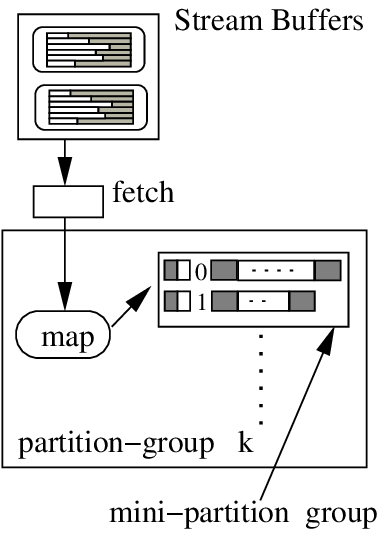}}
\subfigure[]{\label{fig:mpi-fine-tuning}\includegraphics[width=4.5cm]{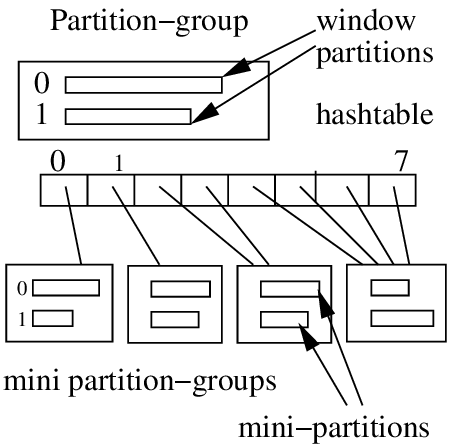}} 
\caption{(a) Processing tuples at a join module, and (b) Fine tuning the window partitions at a slave node based on extendible hashing}
\end{figure}

Each slave node receives  stream tuples from the master at the end of every distribution epoch, and stores the tuples in stream buffers. As discussed earlier, each buffer keeps the tuples for each partition in a separate mini-buffer, so that blocks corresponding to a partition can be retrieved from each mini-buffer without scanning the whole buffer for the stream. The join module fetches a block for a partition of a stream, {\it maps} the tuples to the mini-partition, and keep the tuples in block at the head of the mini window-partition of the stream (Figure~\ref{fig:mpi-join-proc}). When the head block is full or the buffer contains no more blocks for the respective stream partition, the newly added tuples are joined with the mini-partitions from the opposite stream windows. Also, as obvious from the above description, the tuples in a head block may participate in the join when it is not full. To utilize the empty portion of the head block,  we should keep a variable to track the newly added (i.e., {\it fresh}) tuples within a head block. To eliminate duplicate output results, we omit the fresh tuples within the head blocks of the opposite mini window-partitions, as the head blocks will participate in the join when they become full.  While expiring a block from a mini-window, the block is joined with the fresh tuples within the head block of the opposite mini-window; this ensures the completeness of the join results. 

\par The size of a (mini) partition of a stream may grow and shrink depending on the arrival rates of the tuples within the respective partition. As  the size a window partition increases, time needed   join the partition with the tuples from the opposite streams also increases. On the other hand, a small partition size increases the memory overhead. Thus we should adjust the size of the partitions. We adapt the size at the granularity of the partition-groups. We keep the size (in blocks) of each partition-group within a range $[\theta \ldots 2\theta]$. We split a partition-group when its size is above $2\theta$, and try to merge partitions whenever  a partition size falls below $\theta$. We maintain the size of the partition-groups using extendible hashing~\cite{fagin79}, and maintain  one hashtable for each overflowing partition-group (Figure~\ref{fig:mpi-fine-tuning}). We split a partition-group using the extendible hashing, and create multiple mini-partition-groups; each mini-partition-group is pointed to by one or more hashtable entries. 
                      
\par In extendible hashing, the hashtable size (i.e., number of entries in the hashtable) is given by a parameter called the {\it global depth} $d$; the hashtable uses  $d$ least significant bits of some adopted hash function $h(k)$, and has size $2^d$ entries. Each mini-partition-group, that is meant to be a bucket in a hashtable, is assigned a local depth $d'$. The number of hashtable entries pointing to a bucket is given by $2^{d-d'}$; the $d'$ least significant bits of these entries are the same.  When the size of a mini-partition-group, with local depth less than the global depth, is larger than $2\theta$, we split the mini-partition-group by assigning half of the $2^{d-d'}$ entries to each new mini-group and distributing the tuples accordingly; the local depth of the newly created partitions are increased by one. To split a mini-partition-group with the local depth same as the  global depth, the total entries in the hashtable should be doubled first, increasing the global depth; now that the local depth is less than the global depth, the mini-group can be splitted using the same approach as before.

\par  When the size of a mini-partition-group is less than $\theta$, the  bucket is merged with its {\it buddy} bucket, if any, provided the sum of the sizes of two buckets is less than $2\theta$ and the local depths of the two buckets are same. Let $l_{bud}$ be the first entry of the buddy of a bucket with starting  entry $l$; let the local and the global depth be $d'$ $d$, respectively.  
\[ 
l_{bud} =
\begin{cases}  l+2^{d-d'}, \;\mbox{if} \; 2^{d-d'+1}\; \mbox{divides}\; l \\ 
  l-2^{d-d'},\;\mbox{otherwise} 
\end{cases} 
\]   

Within each mini-partition-group, the incoming tuples are joined using a simple Block-Nested Loop join. The tuples within each  window partition should be expired periodically. Thus, the tuples should maintain the temporal order in the stream; this constraint makes any sort-based algorithm infeasible. At first look it appears that tuples in a window partitions can be sorted using a out-of-place storage that stores the sorted tuple IDs, whereas the actual tuples reside in the window partition. Such an approach suffers from both storage overhead and computation overhead due to frequent expiration of the stream tuples.  

\section{Scalability issues} \label{sec:scala}
In this section, we consider the system performance with a large number of processing nodes. Considering the scalability issue, we describe two techniques to control system overheads.   

\subsection{Degree of Declustering}
Determining the appropriate degree of declustering (i.e., number of slave nodes) is an important issue while using intra-operator parallelism. Selecting low degree of declustering can lead to under-utilization of the system (i.e., nodes not actively participating in the processing  waste their idle CPU cycles), and reduce system performance overloading the processing nodes. On the other hand, high degrees of parallelism may under-utilize the processing nodes and increase  communication overhead. The decrease communication overhead, the payload (i.e., number of stream tuples) of the messages  sent, in each distribution epoch,  to every slave node should be as high as possible~\cite{baker99}.    

Setting the upper bound of the degree of parallelism based on bounding the communication cost as in ~\cite{garofal96} is infeasible in the scenario of continuous queries for a number of reasons: firstly, unlike traditional queries, the degree of declustering may vary during execution. Secondly, the communication cost during the execution can't be estimated by a fixed, simple model. Therefore, an adaptive approach is necessary. Moreover, in a multi-process environment, measuring the execution time and communication delay is impossible due to frequent context switching, which is transparent to the processes or threads, by the operating system. Based on these observations, we propose a simple approach to maintain the degree of declustering. 

\par Our approach adjusts the degree of declustering based on the observation of the loads of the processing nodes. The master node decreases the processing nodes when the load of the processing nodes are very low, and increases the active slave nodes when the loads of the processing nodes are high. Our approach keeps the system minimally overloaded by  ensuring at least one supplier in the system. If all the nodes are either {\it neutral} or {\it consumer}, the master node decreases the degree of declustering. This minimizes the underutilization of the active slave nodes. On the other hand, the master node increases the degree of declustering when the number of supplier in the system is greater than $\beta$ times the number of consumers in the system, that is,\[N_{sup} > \beta N_{con}\]
Here, $N_{sup}$ and $N_{con}$ are, respectively,  the number of supplier and consumer in the system; and $\beta$ ($0<\beta<1$) is a granularity parameter. This ensures the proper utilization of the system. 
     
%\begin{figure} 
%\centering
%\includegraphics[width=7cm]{figure/subgroup-comm.eps} 
%\caption{Subgroup communication}\label{fig:mpi-subgroup-comm}
%\end{figure}

\subsection{Sub-group Communication}
The master node distributes the buffered tuples to all the slaves at the start  of every distribution epoch. As the tuples are sent to every node in a serial order, such an approach might increase average  idle time in waiting for the tuples. For example, in a system with $N$ slaves, a slave, in the worst case, should have wait in idle until the master node transmits the tuples to other $N-1$ slaves. To minimize such an overhead, we divide the slave into  a number ($n_g$) of groups, and allow the slaves within each sub-group communicate with the master at a time;  the distribution epoch is divided into $n_g$ slots, and each sub-group receives tuples from the master within a slot assigned to the group.
%The details of the communication is shown in Figure~\ref{fig:mpi-subgroup-comm}. 

Such a communication in sub-group also minimizes  the total storage or memory required to buffer the pending tuples at the master node. Let us consider a stream $S_i$  with a uniform rate $r_i$, and assume that the master distributes an equal number of tuples to every slaves. The total tuples of stream $S_i$ arriving during a distribution epoch $t_d$ is $t_dr_i$. With sub-group communication, the maximum buffer size for the stream at the master node, can be calculated as, 

\begin{eqnarray}
M_{buf}&=&\frac{r_i}{n_g}t_d+ \frac{r_i}{n_g}\left(t_d-\frac{t_d}{n_g}\right)+ \ldots \nonumber \\
 & &  + \frac{r_i}{n_g}\left(t_d-\frac{t_d}{n_g}(n_g-1)\right) \nonumber \\
%&=& \frac{r_i}{n_g} \sum_{k=1}^{n_g}\left[t_d-\frac{t_d}{n_g}(k-1)\right]  \nonumber \\
%&=& \frac{r_it_d}{n_g} \left[n_g-\frac{1}{n_g}\sum_{k=0}^{n_g-1}(k)\right]  \nonumber \\
&=& \frac{r_it_d}{2} \left(1+\frac{1}{n_g}\right)  \nonumber 
\end{eqnarray}
   
From this equation, it is obvious that a for a large value of $n_g$, the maximum buffer size can be reduced almost by a half.
           
\section{Experiments}\label{sec:exp}
This section describes the methodologies for evaluating the performance of load diffusion in executing a  stream join operator, and presents experimental results demonstrating the effectiveness of the proposed load diffusion system based on synchronization of the communication among the nodes.

\subsection{Experimental Methodology}
The following paragraphs describe the major components of our experimental setup: the join techniques we consider, the data sets and the evaluation metrics and the experimental platform.  

\noindent {\bf Join Processing Technique}. As observed earlier in the paper, processing sliding  window joins requires the maintenance of the temporal order of the tuples within a window. So, within each window partition, we apply a simple  Nested Loop Join algorithm. We tune the sizes of the window partitions applying the fine tuning technique described in the paper. 

\noindent {\bf Data Stream Generation}. We evaluate the performance of the load diffusion algorithm, using  synthetic data streams. The streams tuples are generated online during system activity. The stream tuples are generated in real time within the master node using a separate module. The stream generation modules are scheduled during idle period, within each distribution epoch, after the master has already sent the pending tuples to the slaves. 
\par We assume  that tuples within a stream $S_i$  arrive with a Poisson arrival rate $\lambda_i$. The inter-arrival time for each tuple is given by the Poisson process. Each stream tuple has a length of 64 bytes.  The domain of the join attribute $A$ is taken as integers within the range $[0\ldots 10\times 10^6]$. The distribution of the join attribute values for the stream tuples is captured using {\it b-model}~\cite{wang02a}, which is closely related to the "80/20 law" in databases~\cite{gray94}.    

\noindent {\bf Evaluation Metrics}. We evaluate the performance of the system based on the capacity of the system, that is, the maximum stream rates that overload the system.  We provide an indication  of the capacity of the system,   
 measuring a number of parameters: {\it average production delay}, communication time (or overhead), total CPU time, and the window size within a node. We measure the  production delay of an output tuple as the interval elapsed since the arrival of the joining tuple with the more recent timestamp. For example,  if tuples $s_1$ and $s_2$ are the joining tuples of the output tuple $(s_1, s_2)$, where $s_1.t\!>\!s_2.t$ ($s_1$ being the more recent one) and current time is $T_{clock}$, then the delay in producing the output tuple is $(T_{clock}-s_1.t)$. This metric (i.e., average production delay) indicates how quick an output tuple is generated. Thus, this metric also provide an indication of the capacity of the system: when the system is overloaded, the incoming tuples stays a longer period of time in buffer, resulting in a larger production delay. 

\noindent {\bf Experimental Platform.} Unless otherwise stated, the default values used in the experiments are as given in Table 2. We have performed our experiments on cluster of machines connected by a Gigabit Ethernet Switch. Each machine has two Pentium III (coppermine) 930 MHz  processors, 256 KB L2 cache, and 512 MB of main memory. For each experimental setting, we run the system for 20 minutes, and refresh the observed parameters by the elapse of a time of 10 minutes.  At the master size, we provide the level of indirection, while  distributing the tuples, by maintaining 60 partitions; each partitions  stream or window partitions are fine tuned, at the slave side, based on a $\theta$-value of $1.5$MB. We fix the block size to 4KB. We allocate 1MB of memory to buffer the stream tuples. We assume that each processing node has enough memory to hold the window partitions; extension of the system to cope with  memory limited nodes is straightforward, based on the incorporation of the memory occupancy information during partition reorganizations. The  distribution epoch  and the reorganization epoch  are taken as 2 and 4 seconds, respectively. 
\par We implement the join processing algorithm in Java. We use {\it mpiJava}~\cite{baker99}, a pure Java interface to MPI, that uses services of a native MPI implementation. We use LAM/MPI~\cite{burns94} (version 7.0.6) as the underlying native MPI implementation. Unless otherwise stated, the default values used in the experiments are as given in Table 2.  

\begin{table}
\centering
\begin{tabular}{|c|c|l|}
\hline 
Parameter & Defaults & Comment \\ \hline\hline
$W_i (i=1, 2)$ & 10 				& Window length(min)\\ \hline
$\lambda$ & 1500 	& Avg. arrival rate(tuples/sec)\\ \hline  
$b$ & 0.7			& skewness in join attribute  \\
				&		&values(for {\it b-model}\/)\\ \hline
$Th_{con}$ & 0.01 	& Consumer Threshold\\ \hline 
$Th_{sup}$ & 0.5 	& Supply Threshold\\ \hline  	 
$\theta$		&	1.5 & partition tuning parameter (MB) \/)\\ \hline	
				& 4 & Block Size (KB) \\ \hline
$t_{d}$		& 2 & Distribution epoch (sec) \\ \hline
$t_{r}$	& 20 & Reorganization  epoch (sec)\\ \hline   
\end{tabular}
\caption{Default values used in experiments} \label{tab:defaults}
\end{table}

\subsection{Experimental Results}
In this section, we present a series of experimental results for assessing the effectiveness of the proposed hashjoin algorithm. We measure the average delay in generating an output tuple, maximum window sizes in the nodes, processing time, and communication overhead. For each set of experimentation, we run the system for 20 minutes. We start to gather performance data after an startup interval of 10 minutes is elapsed.

\begin{figure}%[h]
\centering
\includegraphics[width=7cm ]{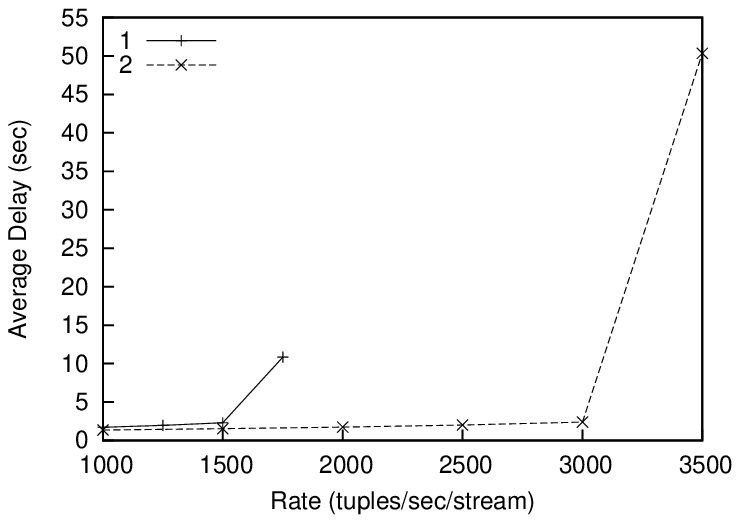}
\caption{ Average delay with varying stream arrival rates }\label{fig:mpi-rate1-vs-avgdelay}
\end{figure}

\begin{figure}%[h]
\centering
\includegraphics[width=7cm ]{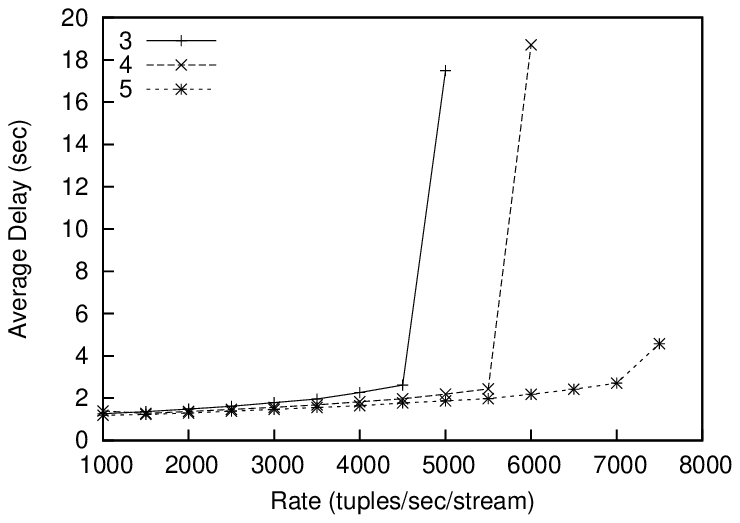}
\caption{ Average delay with varying stream arrival rates }\label{fig:mpi-rate2-vs-avgdelay}
\end{figure}

\par We first present the experimental results with varying stream arrival rates. Figure~\ref{fig:mpi-rate1-vs-avgdelay} and Figure~\ref{fig:mpi-rate2-vs-avgdelay} show the average delay, with varying stream arrival rates, observed in the output tuples.  Each plot in the figures corresponds to different slave population. Given a  number of slave nodes, the average delay increases sharply at a point where the applied load overloads the whole system. Before such a saturation point, the average delay shows very little variations with the increase in stream rates. This is due to the diffusion of the loads from a temporarily overloaded node. From the two figures, we observe that the stream arrival rate which overloads the system increases as more and more slave nodes are added in the system. 

\par
\begin{figure}%[h]
\centering
\includegraphics[width=7cm ]{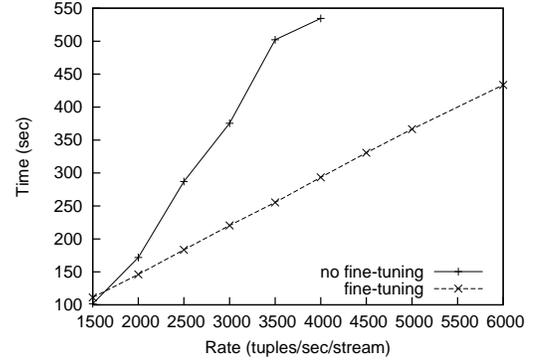}
\caption{ Average processing time (i.e., CPU time) with varying stream arrival rates (total slave nodes=4) }\label{fig:mpi-rate-vs-cpuT}
\end{figure}

\begin{figure}%[h]
\centering
\includegraphics[width=7cm ]{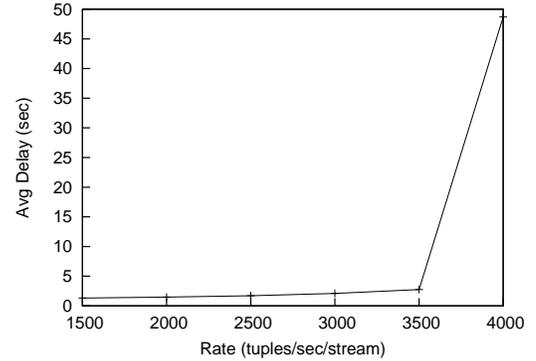}
\caption{ Average delay with varying stream arrival rates (total slave nodes=4) }\label{fig:mpi-rate-vs-NT-delay}
\end{figure}

 The effectiveness of  fine grained partition tuning   at the slave nodes can be observed from the Figure~\ref{fig:mpi-rate-vs-cpuT} and Figure~\ref{fig:mpi-rate-vs-NT-delay}. Figure~\ref{fig:mpi-rate-vs-cpuT}  shows the average CPU times (both with and without fine tuning at the slaves) while processing stream joins with varying arrival rates. Without fine tuning, the average CPU time required to process the joins increases sharply with the increase in the stream rates. As shown in Figure~\ref{fig:mpi-rate-vs-NT-delay}, without fine tuning, the average delay is around 48 sec for a per-stream rate of 4000 tuples/sec. On the other hand, with fine tuning, the average delay for the same system (with 4 slave nodes and for an arrival rate of 4000 tuples/sec) drops to around 2 seconds (cf., Figure~\ref{fig:mpi-rate2-vs-avgdelay}).

\begin{figure}%[h]
\centering
\includegraphics[width=7cm ]{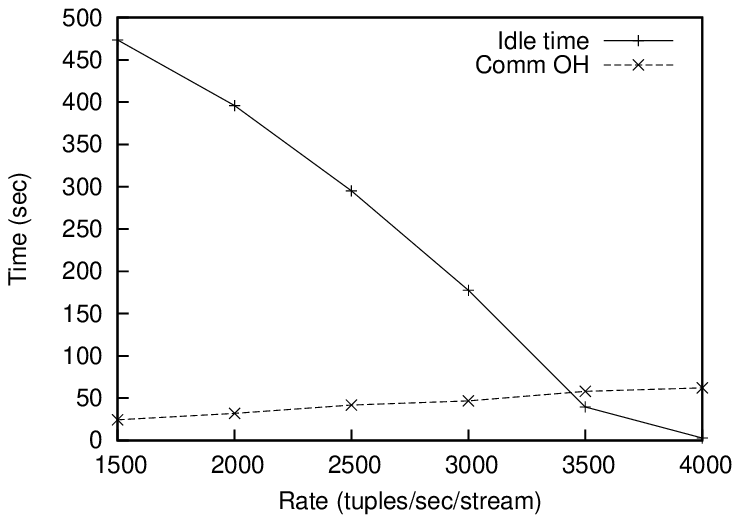}
\caption{ Idle time and communication overhead  with varying stream arrival rates (without fine-grained partition tuning, total slave nodes=4) }\label{fig:mpi-rate-vs-NT-idleTohT}
\end{figure} 

\begin{figure}%[h]
\centering
\includegraphics[width=7cm ]{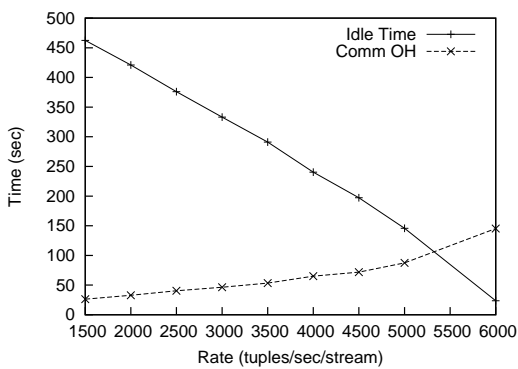}
\caption{ Idle time and communication overhead  with varying stream arrival rates (with fine-grained partition tuning, total slave nodes=4) }\label{fig:mpi-rate-vs-FT-idleTohT}
\end{figure} 
\par  Figure~\ref{fig:mpi-rate-vs-NT-idleTohT} and Figure~\ref{fig:mpi-rate-vs-FT-idleTohT} show the idle time and the communication overhead for the system with and without applying fine grained partition tuning. Without partition tuning, the idle time for the system drops near to zero for an arrival rate of 4000 tuples/sec/stream.  On the other hand, with fine grained partition tuning, the idle time for the slave nodes is near to  zero while the arrival rate is 6000 tuples/sec/stream. With the increase in arrival rates, the sizes of the partitions of each stream window  also increases. Thus scanning the window partitions to join the incoming stream tuples consume higher CPU times  with the increase  in the stream rates. Tuning the partition sizes at the slave nodes significantly lowers the CPU time to process the join. Such a fine  grained tuning incurs no communication overhead as evident from  Figure~\ref{fig:mpi-rate-vs-NT-idleTohT} and Figure~\ref{fig:mpi-rate-vs-FT-idleTohT}. 

\par
\begin{figure}%[h]
\centering
\includegraphics[width=7cm ]{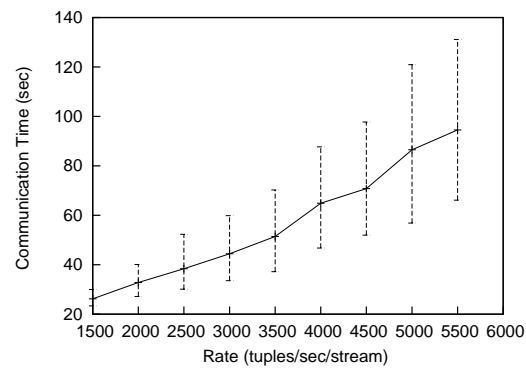}
\caption{ Communication  overhead  with varying stream arrival rates (total slave nodes=4)   }\label{fig:mpi-rate-vs-commOH}
\end{figure}

 Figure~\ref{fig:mpi-rate-vs-commOH} shows the communication overhead across the slave nodes. It shows the minimum, maximum and average communication overhead over all the slave nodes in the system. The communication time is not uniform across the slaves, as the tuples are distributed to the slaves, during a distribution epoch, in a serial order. The divergence in communication overhead  across the slaves increases with an increase in arrival rates. Such divergence across the slave nodes can be minimized by  maintaining a fixed  order while  distributing tuples across the slaves; now, the slave node can delay its connection initiation according to its position in the sequence. 

\begin{figure}%[h]
\centering
\includegraphics[width=7cm ]{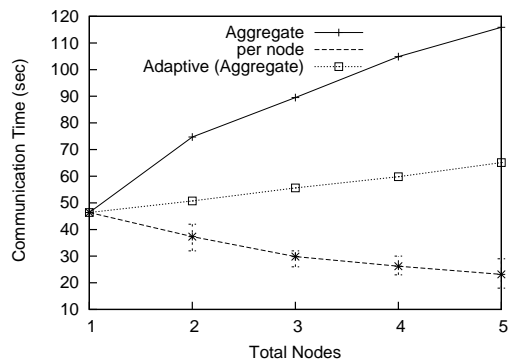}
\caption{ Communication overhead with varying nodes}\label{fig:mpi-node-vs-commOH}
\end{figure} 
 
Figure~\ref{fig:mpi-node-vs-commOH} shows the  communication overhead  with varying slave nodes. The communication time decreases with the increase in  degree  of declustering; however aggregate overhead over all the slaves increases linearly. The figure shows the aggregate overheads with adaptive parallelism. In this case, the system automatically fixes the number of slaves based on instantaneous arrival rates. With adaptive parallelism, the aggregate communication overhead is significantly lower while compared to non-adaptive counterpart (for an stream arrival rate of 1500 tuples/sec/stream). As the overhead increases with stream arrival rates (Figure~\ref{fig:mpi-rate-vs-commOH}), the performance benefit of an adaptive algorithm would be prominent for a high stream arrival rate.So, it is  evident that the degree of declustering of the system should not be increased unless necessary, i.e., all the nodes should operate as close to its processing capacity as possible. 
 
\par
\begin{figure}%[h]
\centering
\includegraphics[width=7cm ]{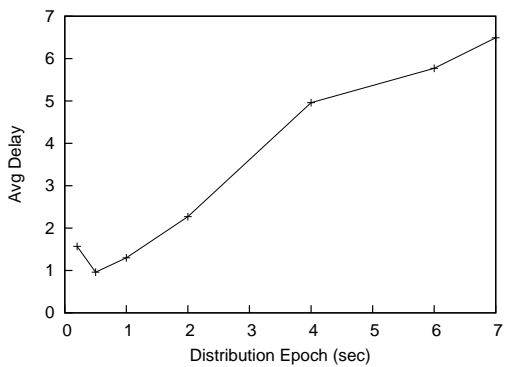}
\caption{Average production delay with varying distribution epochs (total slave nodes=3)}\label{fig:mpi-epochT-vs-delay}
\end{figure}

\begin{figure}%[h]
\centering
\includegraphics[width=7cm ]{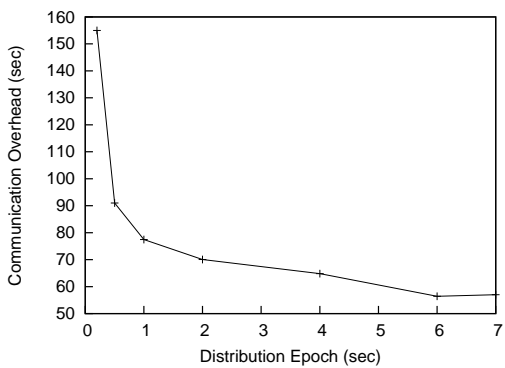}
\caption{Communication overhead with varying distribution epochs (total slave nodes=3)}\label{fig:mpi-epochT-vs-commOH}
\end{figure}

 Figure~\ref{fig:mpi-epochT-vs-delay} and Figure~\ref{fig:mpi-epochT-vs-commOH} shows the average delay of the output tuples  and the average communication overhead across the slave nodes, respectively, with varying distribution epochs. As the distribution epoch decreases, the average delay also decreases due to the decrease in the wait time at the master node.  However, with the decrease in the  distribution epoch, the communication overhead increases (Figure~\ref{fig:mpi-epochT-vs-commOH}) and reaches at point (not shown in the figure)  where the slaves are engaged only in communication, leaving no time for the processor to process the incoming tuples. Thus, for a fixed stream arrival rate, there exists a tradeoff between the distribution epoch and the communication overhead. 

\section{Related Work}\label{sec:relWorks}
Existing relevant work on diffusing loads of a stream join operator can be classified into two categories: recent advancements in continuous query and stream processing systems, and earlier work in parallel query processing. 

\par The STREAM~\cite{motwan03} project focuses on computing approximate results and minimizing the memory requirement of continuous queries over data streams. The Aurora~\cite{carney02} system proposes mechanism to sacrifice result quality, based on  user specified quality-of-service profiles, while sufficient resources to ensure scalability are not available. In contrast, we exploit inexpensive shared-nothing clusters to ensure scalability without sacrificing result accuracy. Reference~\cite{balazin04} proposes a contract-based load management framework migrating workload among  processing nodes based on predefined contracts. The Borealis project proposes a dynamic inter-operator load distribution mechanism by utilizing the operators' load variance coefficients~\cite{xing05}. StreamCloud~\cite{gulisano:streamCloud10} parallelizes a set of stream queries across a number of virtual machines in a cloud. Stormy~\cite{loesing:stormy12} uses techniques from key-value stores and replication to provide a fault-tolerant service for processing streams in a cloud system. In comparison, our work consider intra-operator load distribution for window join queries with large states and high arrival rates.   

\par The Flux operator~\cite{shah03} extends the exchange operator~\cite{graefe90} to support adaptive dataflow partitioning and load balancing while processing stateful operators ( e.g. joins, grouping operators)  over a shared-nothing cluster. The  Flux operator consists of two types  of intermediate operators called Flux-Prod and Flux-Cons. The Flux-Prod operator stores stream tuples (from the sources) into a buffer, and distributes the stream tuples among a number of Flux-Cons operators, that are instantiated in each of the nodes processing the stream queries. The Flux operator provides a framework for partitioning dataflows over a number of nodes; however, it does not consider the sliding window joins over a shared nothing environment. Moreover, while implementing a dataflow operator over a number of processing nodes, maintaining and initiating communication among the nodes, over a reliable socket or TCP connection~\cite{stevens94},  without any prior synchronization or  without any predefined order of data exchange is infeasible, if not impossible. 

\par In reference~\cite{gu07}, the authors first address the issue of intra-operator parallelism while processing a join operator over a number of servers. The paper provides two tuple routing strategies, namely {\it aligned tuple routing} (ATR) and {\it coordinated tuple routing} (CTR), that preserve join accuracy. The ATR assigns a segment of the master stream to a selected node, and changes the assigned node at the end of every segment.  The ATR works for a segment much higher than the sizes of the stream windows. Thus, instead of balancing the loads, this approach circulates the join processing  load  across the nodes: during a segment interval the node assigned with the segment of the master node carries out all the join processing loads, while the remaining slave nodes (assigned with a segment of the slave streams) only forward  the incoming tuples of the slave streams to the respective master node. This approach violates the assumption of  resource limited processing nodes; for example, storing the windows of all the streams in a master node (within a segment of the master stream) is infeasible.  Selecting a small segment length doesn't ameliorate the problem; in this case, when a subsequent segment of the master stream is assigned to a new node,  all the stream windows should be routed to  the new master node, and this happens at the end of every segment. On the other hand, CTR distributes the stream segments across the participating nodes, and maintains a routing path for each stream. The routing path is a sequence of routing hop , and each routing hop $V_i$ is a collection of nodes storing a superset of the $i$-th stream-window in the join order.  In addition to the computational overhead associated with maintaining the routing paths, such an approach incurs high network overhead, as each incoming tuple (from a stream source) or intermediate results (generated by a segment of the intermediate {\it routing hop}) should be forwarded, in a cascading fashion, to every node in the successive {\it routing hop}. 

\par Reference~\cite{ivanov05} presents a stream database system that  provides a generic framework for describing distributed execution strategies as high-level dataflow distribution templates.   The paper implements two partitioning strategies: {\it window split} and {\it window distribute}. Each window partitioning strategy provide a partitioning template (OS-Split or S-Distribute) and a combining SQF (OS-Join or S-Merge). The partitioning schemas  are content-insensitive, and are chosen to meet scientific application requirements. Thus the issue of load imbalance across the partitions doesn't arise  However, the paper considers a homogeneous cluster environment without any  non-query background load. Thus the issues  like dynamic dataflow partitioning and state movement are irrelevant  to such a system. 
  
\par Early work on parallel query processing concentrated on parallelizing individual join operators~\cite{schnei89,imasak02a}. Extensive research has been done on handling  data skew (i.e., a non-uniform distribution of join-attribute values) while parallelizing an operator in a shared-nothing system (e.g., ~\cite{wolf93,hua91,kitsur90,keller91,garofal96,dewit92,harada95,zhou95,bamha99,xu08}). These algorithms split the persistent relations in a distribution or splitting phase, and balances the loads by properly assigning the partitions or buckets across the nodes. Such a holistic approach based on the complete knowledge of data distribution in  static data sets is infeasible in the streaming scenario.  In~\cite{dewit92} and~\cite{rahm95}, the authors  describe how to leverage current CPU utilization, memory usage, and I/O load to select processing  nodes and determine degree of declustering for hash joins. Reference~\cite{raman05} presents a method of parallel query processing that uses non-dedicated, heterogeneous computers. This approach assumes a shared stored system (e.g., Storage Area Network~\cite{gibson00}) to stores input data, and is not relevant to online stream processing as considered in our work.  The hashjoin algorithms used by all the previous schemes  partition data at a single time point (i.e., between the {\it build} and {\it probe} phases), and they do not consider continual, on-the-fly repartitioning of the join operator during execution.

\section{Conclusion and Future Work}\label{sec:conclusion}
In this paper, we present a new technique  to achieve fine-grained, intra-query parallelism while processing a sliding window  operator over a cluster of processing nodes. The proposed algorithm balances the join processing  loads over a shared-nothing cluster. We analyze the issues in scaling the intra-query parallelism over a large number of nodes in a multi-query, multi-user environment, and propose techniques to dynamically maintain the degree of declustering, optimizing the processing and communication overheads of the system. Our experimental results demonstrate the effective of the algorithm. The work on parallelizing the stream joins  can be extended in a few directions. Most importantly, deploying the prototype over a large number of  processing nodes and dynamically tuning various performance parameters (i.e., group size and distribution epoch) is an interesting future work. Incorporating disk-I/O at the local nodes is another topic for future work.

%\nocite{xu08,wolf91,mehta95,bamha99, raman05, gu05,shah03, imasak02a, zhou95,dewit92,harada95,keller91,graefe90,garofal96, ganguly92, balazin04,xing05,kitsur90,stoneb86,hua91}
 
%ACKNOWLEDGMENTS are optional
%\section{Acknowledgments}
%
% The following two commands are all you need in the
% initial runs of your .tex file to
% produce the bibliography for the citations in your paper.
%\bibliographystyle{latex8}
\bibliographystyle{IEEEtran}
\bibliography{data-stream-biblio}  % sigproc.bib is the name of the Bibliography in this case

% You must have a proper ".bib" file
%  and remember to run:
% latex bibtex latex latex  
%to resolve all references
%
% ACM needs 'a single self-contained file'!
%
%APPENDICES are optional
%\balancecolumns

%\appendix
%Appendix A

\end{document}